\documentclass[12pt,preprint]{aastex}
\usepackage{graphicx}
\begin{document}
\def\la{\mathrel{\hbox{\rlap{\hbox{\lower4pt\hbox{$\sim$}}}\hbox{$<$}}}}
\def\ga{\mathrel{\hbox{\rlap{\hbox{\lower4pt\hbox{$\sim$}}}\hbox{$>$}}}}
\def\lam{$\lambda$}
\def\kms{km~s$^{-1}$}
\def\vphot{$v_{\rm phot}$}
\def\ang{~\AA}
\def\syn{{\bf Synow}}
\def\dm15{{$\Delta$}$m_{15}$}
\def\rsi{$R$(Si~II)}
\def\v10{$V_{10}$(Si~II)}
\def\wsi{$W_lambda$(Si~II)}
\def\vdot{\.v(Si~II)}
\def\575{$W(5750)$}
\def\610{$W(6100)$}
\def\tex{$T_{\rm exc}$}
\def\ve{$v_{\rm e}$}

\title {Comparative Direct Analysis of Type~Ia Supernova Spectra.
II. Maximum Light}

\author {David Branch\altaffilmark{1}, Leeann Chau
Dang\altaffilmark{1},\altaffilmark{2}, Nicholas Hall\altaffilmark{1},
Wesley Ketchum\altaffilmark{1}, Mercy Melakayil\altaffilmark{1}, Jerod
Parrent\altaffilmark{1}, M.~A. Troxel\altaffilmark{1},
D.~Casebeer\altaffilmark{1}, David~J. Jeffery\altaffilmark{1}, \&
E.~Baron\altaffilmark{1}}

\altaffiltext{1} {Homer L. Dodge Department of Physics and Astronomy,
University of Oklahoma, Norman,~OK 73019; e-mail: branch@nhn.ou.edu}

\altaffiltext{2} {Department of Astronomy, Whitman College, Walla
  Walla,~WA 99362}

\begin{abstract}

A comparative study of near--maximum--light optical spectra of 24
Type~Ia supernovae (SNe~Ia) is presented.  The spectra are quantified
in two ways, and assigned to four groups.  Seven ``core--normal''
SNe~Ia have very similar spectra, except for strong high--velocity
Ca~II absorption in SN~2001el.  Seven SNe~Ia are assigned to a
``broad--line'' group, the most extreme of which is SN~1984A.  Five
SNe~Ia, including SN~1991bg, are assigned to a ``cool'' group. Five
SNe~Ia, including SN~1991T, are assigned to a ``shallow--silicon''
group.  Comparisons with \syn\ synthetic spectra provide a basis for
discussion of line identifications, and an internally consistent
quantification of the maximum--light spectroscopic diversity among
SNe~Ia.  The extent to which SN~Ia maximum--light spectra appear to
have a continuous distribution of properties, rather than consisting
of discrete subtypes, is discussed.

\end{abstract}

\keywords{supernovae: general}

\section{INTRODUCTION}

In the first paper of this series (Branch et~al. 2005; hereafter
Paper~I), the parameterized supernova synthetic--spectrum code \syn\
was used to fit and interpret a time series of spectra of the
well--observed, spectroscopically normal, Type~Ia (SN~Ia) SN~1994D.
In this paper we concentrate on near--maximum--light spectra of 24
SNe~Ia.  Maximum--light spectra are of particular interest because
they are the ones most likely to be obtained, especially for
high--redshift SNe~Ia.  In \S2 the spectra are quantified and arranged
to illustrate relationships between them.  In \S3 to \S6 four groups
of spectra are discussed, and selected comparisons with \syn\
synthetic spectra are shown.  Results are discussed in \S7.

As in Paper~I, we confine our attention to optical spectra, from the
Ca~II H\&K feature in the blue ($\sim3700$\ang) to the Ca~II infrared
triplet (Ca~II IR3) in the red ($\sim9000$\ang).  The 24 spectra
selected for study were obtained within three days of $B$--band
maximum light [which occurs about 18~days after explosion (Vacca \&
Leibungut (1996)], a time interval short enough that spectroscopic
evolution is not strong, yet long enough to allow some events of
special interest to be included.  All spectra have been corrected for
the redshifts of the parent galaxies.  The original spectra have a
range of slopes, owing to intrinsic differences, differences in the
amount of reddening by interstellar dust, and observational error.  In
this paper, because we are interested only in the spectral features,
and not in the underlying continuum slopes, the spectra are tilted by
multiplying the flux by $\lambda^\alpha$, with $\alpha$ chosen to make
the flux peaks near 4600\ang\ and 6300\ang\ equally high.  The tilting
makes it much easier to compare the spectral features.  The events of
the sample are listed in Table~1, together with parameters to be
discussed below.

\section{ARRANGING THE SPECTRA}

Various ways of arranging the spectra according to measurements of
selected features were explored.  The arrangement shown in Figure~1
appears to be useful.  The \575\ parameter is the equivalent width of
the absorption near 5750\ang, usually attributed to Si~II \lam5972,
and \610\ is that of the stronger absorption near 6100\ang, produced
by Si~II \lam6355.  (Because of the presence of emission features,
these are equivalent--width--{\sl like} quantities rather than true
equivalent widths.)  The wavelength limits of integration are chosen
by eye\footnote{A precisely defined method, such as used by S.~Bongard
et~al. (in preparation) will be required when strictly reproducible
results are needed.} in the same way as they are for measuring the
\rsi\ parameter, which is the ratio of the fractional depths of these
two absorptions (Nugent et~al. 1995), and the flux used as the divisor
of the intregral is the mean of the fluxes at the wavelength limits.
The ratio of \575\ to \610\ is correlated with, but not equal to,
\rsi.  For our present purpose of arranging the spectra, the
measurement errors are not important, even though the fractional error
of \575\ is large when \575\ is small.

Plotting \575\ against \610\ illustrates the relationships according
to only two spectral features.  To quantify the relationships over a
broader wavelength range, the spectra were interpolated onto a
standard wavelength grid of 2\ang\ separation, and for each pair of
spectra the $\chi^2$--like figure of merit formed from the logarithmic
flux differences between 3700 and 6500\ang\ (or as much of this
interval as possible) was minimized by scaling the flux and varying
the tilt (Appendix~A).  The results identify the nearest neighbor of
each event.  In Figure~1, a single--headed arrow points to a nearest
neighbor (e.g., SN~2002bf is the nearest neighbor of SN~1984A) and a
double--headed arrow connects events that are mutual nearest
neighbors.

In \S3 to \S6 four groups of spectra, denoted by different symbols in
Figure~1, are considered.  The group assignments were made on the
basis of Figure~1, as well as on the appearance (depth, width, and
shape) of the 6100\ang\ absorption feature. Some comparisons with
synthetic spectra are shown, and line identifications are discussed.

\section{CORE--NORMALS}

Figure~2 shows spectra of seven SNe~Ia that are tightly clustered in
Figure~1 and have similar 6100\ang\ absorptions, not only in
equivalent width but also in shape.  Other features, even weak ones,
also are similar.  These will be referred to as ``core--normal''
SNe~Ia.  The only conspicuous difference among them is the strong
high--velocity Ca~II IR3 absorption near 8000\ang\ in SN~2001el.

In Branch et~al. (2003) and in Paper~I, \syn\ fits to the spectra of
two core--normals, SN~1998aq and SN~1994D, were presented and
discussed.  Descriptions of the \syn\ code and its use can be found in
these two papers and references therein.  For the present work, we
adopted precepts for fitting spectra of core--normals that differ
mildly from those of previous papers.  For photospheric--velocity (PV)
ions the line optical depth $e$--folding velocity, \ve, is 1000~\kms.
For most detached high--velocity (HV) ions, \ve\ = 2000~\kms.
(HV~Ca~II in SN~2001el is an exception.) The excitation temperature,
\tex, is 10,000~K.  The ions used are PV~O~I, Mg~II, Si~II, Si~III,
S~II, Ca~II, Fe~III, Co~II, and Ni~II, HV~Ca~II, and HV~Fe~II.  In
Paper~I, Na~I also was used, but the identification (at maximum light)
was not considered to be definite.  In this paper Na~I is not used for
core--normals, on the grounds that if Na~I were present in
core--normals then it should be much stronger in cooler
SN~1991bg--like events, which is not the case.  Lines of C~II also are
not used, because if present at maximum light they are weak and
difficult to confirm [although they appear to be present at maximum
light in at least SN~1998aq (Branch et~al. 2003), and will be used in
Paper~III (in preparation) on premaximum--light spectra].  In Paper~I,
HV~Fe~II was not used in near--maximum--light spectra.  In this paper
HV~Fe~II is used\footnote{In the course of this work the Fe~II
log~{\sl gf} values of the Kurucz (1993) line list, on which {\bf
Synow} relies, were compared with those derived by Phillips (1979)
from the solar spectrum. Among the strong lines two significant
discrepancies were found, and the Phillips values of log~{\sl
gf}(\lam5169)=$-1.66$ and log~{\sl gf}(\lam4549)=$-2.67$ were
adopted.} because we now have stronger evidence for its presence, as
discussed below. The \syn\ fitting parameters for the core--normals
are listed in Table~2.  

In Figure~3 the spectrum of SN~1994D is compared with two synthetic
spectra.  As usual, the blackbody continuum of \syn\ makes the
synthetic spectrum too high from about 6500\ang\ to 7800\ang.  (The
continuum of SNe~Ia cannot be modeled as a single blackbody curve, but
an attempt to model the exact continuum shape at the \syn\ level of
analysis would be an unrewarding fitting exercise.)  One synthetic
spectrum includes HV~Fe~II, detached at 16,000~\kms, and the other
does not.  The main effect of HV~Fe~II is produced by two lines of
multiplet~42, \lam5018 and \lam4924.  (The third line of the
multiplet, \lam5169, is in this case overwhelmed by Si~II absorption.)
In SN~1994D and most other events of the sample, an absorption near
4430\ang, which we believe to be produced by Si~III \lam4550, is not
well fitted because the Si~III synthetic absorption is too blue.  The
fit could be improved by lowering \vphot\ and slightly detaching all
ions other than Si~III, but this would not help us to understand the
discrepancy so we have refrained from doing it.  Because the synthetic
spectra usually are too high in the red, and because of the presence
in some spectra of the telluric absorption near 7600\ang, it is
difficult to know how to fit the absorption produced by O~I \lam7773
(and Mg~II \lam7890), so little significance should be attached to the
fitting parameters used for O~I.

In Figure~4 the spectrum of SN~2001el is compared with a synthetic
spectrum in which HV~Ca~II has a high optical depth ($\tau=12$,
compared to $\tau=2$ to 5 for other core--normals), a high value of
\ve\ = 12,000~\kms, and a maximum velocity of 30,000~\kms.  The high
value of \ve\ makes the absorption flat and extend to the maximum
velocity.  This fit is not unique, but it demonstrates that HV~Ca~II
can account for the feature.  Polarization observations indicate that
asymmetry, not taken into account in \syn, affects HV~Ca~II line
formation in SN~2001el (Wang et al. 2003, Kasen et~al. 2003). The best
fit for SN~2001el is obtained with a HV~Fe~II optical depth that is
the highest among the core--normals.  Since SN~2001el has strong
HV~Ca~II absorption, this is regarded as support for the HV~Fe~II
identification.  The normality of all features {\sl not} affected by
HV~Ca~II and HV~Fe~II indicates that the PV spectrum of SN~2001el is
very much like those of the other core--normals.

Most of the spectral features of the core--normals have established
line identifications.  Ions whose presence is considered to be
definite are PV Si~II, S~II, Ca~II, O~I, Mg~II, Fe~III, and HV Ca~II.
Co~II and Si~III are probable, while Ni~II is possible, but difficult
to establish.  We will return to the issue of HV~Fe~II below.  The
issue of C~II will be addressed in Paper~III (in preparation) on
premaximum spectra.  A few of the weak features of the core--normals
are not reproduced by the \syn\ spectra.  Possible identifications of
the weak absorptions near 5500\ang\ and 5150\ang\ (seen most clearly
in Figure~3) are Si~III \lam5740 and Co~II \lam5311, respectively.
The weak absorptions near 4150\ang\ and 4550\ang\ (see Figure~2)
remain unidentified (but see \S7 on the 4550\ang\ absorption).

\section{BROAD--LINE SNe~Ia} 

Figure~5 shows the spectrum of the core--normal SN~1994D (for
comparison) and the spectra of seven SNe~Ia that are spectroscopically
normal in the traditional sense that their strongest spectral features
are the usual ones, rather than the conspicuous Fe~III features of
SN~1991T or the blue Ti~II trough of SN~1991bg.  However, these seven
``broad--line'' SNe~Ia have 6100\ang\ absorptions that are broader and
deeper than those of the core--normals, which puts them to the right
of the core--normals in Figure~1.  The spectra of SN~1992A and
SN~1981B are not very different from core--normal, but their lines are
broader, e.g., notice the lack of resolution in the absorption complex
from 4600\ang\ to 5100\ang.  Therefore, we regard SN~1992A and
SN~1981B as small steps from the core in the direction of SN~1984A.
The spectra of Figure~5 do not appear to comprise a simple
one--dimensional sequence.  In SN~2002bf the absorption near 7500\ang\
is broad and shallow.  SN~2001ay has a weak Ca~II IR3 absorption and
steep pieces of spectra near 4000\ang\ and 4200\ang.  [Extreme
photometric peculiarities of SN~2001ay are discussed by Howell \&
Nugent (2004).]

The fitting parameters for the broad--line SNe~Ia are in Table~3.  The
ions are the same as for the core--normals. Reasonable fits were
obtained with \vphot\ = 12,000~\kms, a typical value for core-normals,
and higher values of \ve: 2000~\kms\ for both PV and HV ions, or
higher as indicated in Table~3.  For consistency with the modeling of
the core--normals, PV and HV components were used for Ca~II, but these
broad Ca~II features also can be fitted as single high--\ve\
components, so the Ca~II fitting parameters are far from uniquely
determined.

In Figure~6 the spectrum of the moderately broad--line SN~2002bo is
compared with two synthetic spectra.  HV~Fe~II, with a high optical
depth of 1.5, affects the synthetic spectrum in several places.  In
Figure~7 SN~1984A is compared with synthetic spectra in which
Si~II has \ve\ = 5000~\kms\ and a maximum velocity of 25,000~\kms.
The quality of these two fits is about the same as for the
core--normals.

Line identifications in the broads are the same as in the
core--normals.  Given the non--uniqueness of \syn\ fits (e.g., to an
extent, lowering \ve\ can be compensated by raising the optical
depth), most of the differences between the optical depths used for
the broads and the core--normals are not necessarily significant.

\section{COOL SNe~Ia}

Figure~8 shows spectra of the core--normal SN~1990N and five ``cool''
SNe~Ia, four of which have a conspicuous absorption trough from about
4000 to 4400\ang, due in part to PV Ti~II (Filippenko et~al. 1992a;
Mazzali et~al. 1997; Garnavich et~al. 2004).  SN~1989B, on the other
hand, is spectroscopically normal in the traditional sense, and not
far from core--normal, but its 6100\ang\ absorption is deeper than in
the core--normals.  Its \610\ value is similar to that of SN~1981B,
which is one of the broad--line events of \S4, but the features of
SN~1989B are better resolved than those of SN~1981B.  Notice, for
example, the well resolved region between 4600\ang\ and 5100\ang\ in
SN~1989B.  The 6100\ang\ absorption of SN~1989B is almost identical to
that of SN~1986G.  For these reasons, we regard SN~1989B as a small
step from the core in the direction of SN~1991bg--like spectra, and it
is included with the cools even though it does not have the blue
absorption trough.

The fitting parameters for the cool SNe~Ia are in Table~4.  SN~1989B
can be fitted with parameters that differ only mildly from those of
the core--normals.  For the others of Figure~8, the PV excitation
temperature is lowered from 10,000 to 7000~K in view of the
development of Ti~II lines, which require low temperature.  In
addition to Ti~II, five more ions, Na~I, Mg~I, Ca~I, Sc~II, and Cr~II,
are introduced.  Garnavich et~al. (2004) used Mg~I and Ca~I in their
study of SN~1999by, and Sc~II also is plausible at low temperature.
(Sc~II lines appear in SNe~II when they become sufficiently cool.)
Si~III and Fe~III require high temperature so they are not used for
the cools.

The spectrum of SN~1986G cannot be well fitted with the precepts that
were adopted for the core--normals; when the 6100\ang\ absorption is
fitted, the 5750\ang\ absorption, which is strong in the observed
spectrum, hardly appears in the synthetic spectrum.  A plausible
solution is to impose a maximum velocity on Si~II.  In Figure~9 the
spectrum of SN~1986G is compared with a synthetic spectrum in which
Si~II has a maximum velocity of 15,000~\kms\ and a high optical depth
of 100. Owing to the maximum velocity, the 6100\ang\ absorption
saturates, and owing to the high optical depth, the 5750\ang\
absorption appears.

It has been suggested (Garnavich et~al. 2004) that Ti~II may
contribute to the 5750\ang\ absorption in SN~1991bg--like spectra, but
we find that when the Ti~II optical depth is chosen to fit the blue
trough, Ti~II makes no significant contribution to the 5750\ang\
absorption.  (If the Ti~II optical depth is raised, or an improbably
high value of \tex\ is used, an absorption due to Ti~II lines does
appear near, but not near enough, to the 5750\ang\ absorption.)  On
the basis of detailed spectrum calculations with the {\tt PHOENIX}
code (Baron et~al. 2003 and references therein), S.~Bongard et~al. (in
preparation) also conclude that Ti~II makes no contribution to the
5750\ang\ absorption. Again, a plausible solution is to impose a
maximum velocity on Si~II.  This is shown in Figure~10 for SN~1991bg;
in the synthetic spectrum the Si~II maximum velocity is 14,000~\kms\
and the optical depth is 200.

In view of the large number of ions that may be present, line
identifications in the cools (other than SN~1989B) are a challenge.
Si~II, Ca~II, Mg~II, Ca~II, and Ti~II can be regarded as definite.
(Contrary to previous opinion, S~II cannot, because Sc~II has lines
that could be wrongly attributed to S~II.)  As shown in Figures~9 and
10, Cr~II, Sc~II, Mg~I, Ca~I, Na~I, Co~II, and Ni~II can be combined
to obtain reasonable fits; all of these ions are plausible, but none
are definite.

\section{SHALLOW--SILICON SNe~Ia}

Figure~11 shows spectra of the core--normal SN~1994ae and five
``shallow--silicon'' SNe~Ia that have smaller values of \575\ and
\610, thus occupying the lower left corner of Figure~1.  Despite their
shallow Si~II absorptions, SN~1999ee and SN~1999aw are
spectroscopically normal in the traditional sense.  SN~1991T and
SN~2002cx are spectroscopically peculiar because Fe~III produces the
strongest features (Filippenko et~al. 1992b; Ruiz--Lapuente
et~al. 1992; Jeffery et~al. 1992; Mazzali, Danziger, \& Turatto 1995;
Fisher et~al. 1999).

In Figure~12 the spectrum of SN~1999ee is compared with that of the
core--normal SN~2001el.  SN~1999ee has weaker Si~II and S~II features,
but its HV~Ca~II is almost as strong as that of SN~2001el (Mazzali
et~al. 2005a), and the features attributed to HV~Fe~II are about as
strong as in SN~2001el.  This is additional support for the HV~Fe~II
identification.

The fitting parameters for the shallow--silicon SNe~Ia are in Table~5.
The value of \ve\ is 1000~\kms\ except where indicated otherwise in
the Table.  In Figure~13 the spectrum of SN~1999ee is compared to a
synthetic spectrum.  The ions used are the same as in the core
normals, but with high values of \ve\ = 4000~\kms\ for HV~Fe~II and
6000~\kms\ for O~I and HV~Ca~II.

Figure~11 shows that the spectra of SN~1999ee, SN~1999aw, and
SN~2000cx have many similarities at wavelengths longer than about
4300\ang.  However, compared to SN~1999ee and SN~1999aw, SN~2000cx is
strongly depressed from about 3800\ang\ to 4200\ang.  In Figure~14 the
spectrum of SN~2000cx is compared to a synthetic spectrum in which
Si~II is mildly detached at 15,000~\kms.  As in Branch et~al. (2004a),
HV~Ti~II is used to provide blocking in the blue, and to account for
the sharp flux peak near 4150\ang.  Branch et~al. were unable to
suggest an identification for the distinct absorption feature near
4530\ang, except for HV~H$\beta$, which would be surprising.  However,
as shown in Figure~15, we now realize that HV~Cr~II can account for
this feature.  If the HV~Ti~II identification is correct, then
HV~Cr~II, at practically the same detachment velocity, is a more
plausible identification than H$\beta$.  In the synthetic spectrum of
Figure~14 HV V~II also is introduced, but its main effect is in a
crowded spectral region and the identification is not definite.

SN~1991T may be a continuation of the sequence shown in Figure~11;
e.g., it bears some resemblance to SN~1999aw, although its Ca~II and
Si~II features are weaker.  As further discussed in \S7, even though
SN~1991T and SN~2002cx are mutual nearest neighbors spectroscopically,
in several respects they are quite different. It may be that the only
thing these two have in common is a high temperature.

\section{DISCUSSION}

The high degree of spectral homogeneity among the core--normals
suggests that they result from a standard, repeatable SN~Ia explosion
mechanism that does not produce spatially large composition
inhomogeneities near the characteristic photospheric velocity of
12,000~\kms\ (Thomas et~al. 2004).  The strong HV~Ca~II of SN~2001el
may involve a spatially large structure near 20,000~\kms\ (Kasen
et~al. 2003; Kasen \& Plewa 2005).  On the other hand, the features
attributed to HV~Fe~II are only mildly enhanced in SN~2001el and
rather homogeneous in the others, suggesting that weak HV~Fe~II is
``natural'' in SNe~Ia, in the sense that it does not require unusual
high--velocity structures (H\"oflich, Wheeler, \& Thielemann 1998,
Lentz et~al. 2001b).  Weak HV~Ca~II also is ubiquitous (Mazzali et
~al. 2005b).

Based on previous work (Branch 1987; Benetti et~al. 2004) we expected
to fit the broad--line SNe~Ia with high values of \vphot, but better
fits were obtained by holding \vphot\ constant at 12,000~\kms, a
typical value for the core--normals, and using increased \ve\ values.
This may be a consequence of nuclear burning extending to higher
velocity in the broads than in the core--normals (Lentz et~al. 2001a;
Benetti et al. 2004; Stehle et~al. 2005).

Although the composition structures of the cools surely differ from
those of the core--normals (e.g., the low luminosity requires a low
mass of $^{56}$Ni), the immediate reason for the spectroscopic
differences appears to be the lower temperature.  The blue trough in
cools, usually referred to as the Ti~II trough, is produced not only
by Ti~II, but also by other ions that appear at low temperature.
Ti~II is not significant for the 5750\ang\ absorption.  The 5750\ang\
absorption of the cools can be accounted for in terms of a high Si~II
optical depth together with a maximum Si~II velocity.  This is a
plausible way to simultaneously fit these two absorptions while
explaining why \rsi\ and the ratio of \575\ to \610\ go the wrong way
with temperature (i.e., increasing with falling temperature),
considering that the lower level of \lam5972 is the upper level of
\lam6355. If correct, this maximum velocity would indicate that in the
cools nuclear burning extends only to unusually low velocities, near
15,000~\kms.

In most of the shallow--silicon SNe~Ia the spectral features appear to
be produced by the same ions as in the core--normals, although the
optical depths are quite different, especially in SN~1991T and
SN~2002cx.  Most of the differences from core--normals may be caused
by a higher temperature.  Like SN~2001el among the core--normals,
SN~1999ee provides support for the HV~Fe~II identification.  The cause
of the depression of the spectra of SN~2000cx in the blue from about
3800\ang\ to 4200\ang\ appears to be extra HV line blocking.  If HV
Ti~II is present, then HV~Cr~II is a more plausible identification
than HV H$\beta$ (Branch et~al. 2004a) for the 4530\ang\ absorption
in SN~2000cx.  As noted in \S3, a less distinct absorption appears
near 4550\ang\ even in the core--normals; HV Cr~II is a possibility.

One of the most interesting issues that can be addressed by studies of
this kind is whether SNe~Ia have a continuous distribution of
properties, or consist of discrete subgroups.  On the whole, we have a
sense of continuity.  SN~1992A and SN~1981B may be small steps from
the core in the direction of SN~1984A; SN~1989B may be a step in the
direction of SN~1986G and SN~1991bg; SN~1999ee may be a step in the
direction of SN~1991T.  If the distribution is continuous, then
exactly how the boundaries of the core--normal group are defined is
not important.

The cools stand apart from the others in Figure~1, but this in itself
does not establish that they differ in kind from the others.  As noted
by Hatano et~al. (2002) and H\"oflich et~al. (2002), there is a
temperature threshold below which, owing to abrupt changes in key
ionization ratios, line optical depths change abruptly (Hatano
et~al. 1999).  Therefore, the temperature interval required to go from
SN~1989B through SN~1986G to SN~1991bg--like events may be narrow.
SN~1991bg--like events also are quite subluminous, but especially in
the $B$ band, which is strongly blocked by the blue trough.
Similarly, there is a temperature threshold above which Fe~III
features become conspicuous (Hatano et~al. 2002).  Most SNe~Ia may be
variations on the basic core--normal theme, with the diversity
reflecting differences of degree, rather than differences of kind.

Evidence recently has been presented that in terms of stellar
population age there are two SN~Ia groups of roughly equal size, one
from a young ($\sim10^8$ years) population and another from another
from an older ($\sim3$~Gyr) population (Mannucci et~al. 2005a;
Scannapieco \& Bildsten 2005; Mannucci, Della Valle, \& Panagia
2005b).  However, we see no basis for splitting the SNe~Ia of our
sample into two groups of roughly equal size.  It may be that there
are two different binary--star evolution paths to SNe~Ia, one
requiring a short time and the other a long time, but with the two
paths producing similar outcomes, i.e., carbon--oxygen white dwarfs
that explode as they approach the Chandrasekhar mass.

The one event of our sample that does seem to be different in kind is
SN~2002cx, which we have placed in the shallow--silicon group. Like
SN~1991T it had conspicuous Fe~III lines, but unlike SN~1991T it did
not have a slow light--curve decline rate and it was subluminous
(Li~et~al. 2003).  Its late--time spectra were unlike those of normal
SNe~Ia, being dominated by very low velocity ($\sim$700~\kms)
permitted lines of Fe~II, Ca~II, Na~I, and perhaps O~I.  Several other
SN~2002cx--like events have been discovered (S.~Jha et~al., in
preparation).  On the basis of the first--season observations by
Li~et~al. (2003), Branch et~al. (2004b) suggested that SN~2002cx may
have been a pure deflagration while most other SNe~Ia, even
SN~1991bg--likes (H\"oflich et~al. 2002), may be delayed detonations
or gravitationally confined detonations (Plewa, Calder, \& Lamb 2004;
Kasen \& Plewa 2005). However, the late--time spectra of SN~2002cx
(S.~Jha et~al., in preparation) do not resemble the spectra calculated
by Kozma et~al. (2005) for one particular 3--D deflagration model, so
the nature of SN~2002cx remains uncertain.

The normality of PV features in events such as the core--normal
SN~2001el and the shallow--silicon SN~2000cx, both of which have
strong HV features, suggests that HV and PV are somewhat independent
(with HV~Fe~II and weak HV~Ca~II being ubiquitous and natural, as
discussed above).  An interesting issue for further study is how much
of the photometric diversity of SNe~Ia is caused by HV features (which
may not correlate with luminosity).  Another issue is whether the
strong HV blue line blocking of SN~2000cx occurs only in the
shallow--silicon group.  It is clear that strong HV~Ca~II IR3
absorption occurs not only in the core--normal SN~2001el but also in
the shallows.

Our study is complementary to that of Benetti et~al. (2005; hereafter
Be05), who did not consider the strenths of spectral features. Sixteen
events of our sample were included in the study of Be05, who assigned
SNe~Ia to three groups on the basis of absolute magnitude ($M_B$);
light--curve decline rate (\dm15; Phillips 1993); the blueshift of the
6100\ang\ absorption 10 days after maximum [\v10; Branch \&
van~den~Bergh 1993]; the rate at which the 6100\ang\ absorption drifts
to the red [\vdot; Be05]; and \rsi.  With a few borderline exceptions
(SN~1989B and SN~1992A), our cools corresponds to the Be05 FAINT
group, which is to be expected since both temperature and luminosity
are controlled mainly by the $^{56}$Ni mass.  Our broad--line group
corresponds to the Be05 HVG (high temporal--velocity gradient) group.
This also makes sense because broad 6100\ang\ absorption requires high
Si~II optical depth over a substantial velocity range, which makes it
possible for the absorption minimum to shift appreciably with time.
The broads appear to have thicker silicon layers than the
core--normals.  Our core normals and shallows correspond to the Be05
LVG (low temporal--velocity gradient) group.  The lower velocity range
over which Si~II has a high optical depth permits only a smaller shift
in the absorption minimum with time.

Polarization observations (Wang et~al. 2001; Howell et~al. 2001; Wang
et~al. 2003; Leonard et~al. 2005) show that some spectral features are
affected by asymmetric structures.  A straightforward way to rule out
asymmetry as the sole cause of the differences between groups of
SNe~Ia would be to demonstrate a correlation with parent galaxy type
(Be05) or other indicators of the parent--galaxy population (Gallagher
et~al. 2005).  If asymmetry were the sole cause of SN~Ia differences,
SN~Ia subtypes would not correlate with galaxy types.  But, in fact,
there are correlations and emerging correlations.  SN~1991bg--like
events tend to occur in older populations (Howell 2001).  In the Be05
sample the mean host galaxy morphological type (the T~type) was found
to increase from the FAINT to HVG to LVG groups.  Similarly, for our
sample the mean T~type increases from cools to core--normals to broads
to shallows, but the statistics are too poor to allow definite
conclusions.

The expected rapid increase in the number of well observed SNe~Ia in
the near future (e.g., Wood--Vasey et~al. 2004) will allow the
relationships among SNe~Ia to be clarified.  We expect that our
Paper~III (in preparation), on premaximum spectra, also will clarify
the view presented here.

We are grateful to all observers who have provided spectra.  This work
has been supported by NSF grants AST-0204771 and AST-0506028, and NASA
LTSA grant NNG04GD36G.

\appendix

\section{Measuring Spectral Difference}

    In order to quantify difference in spectral features we use the
scaled--tilted spectrum expressed by flux values $\tilde y_{\ell i}$
for wavelengths $\lambda_{i}$ which are parameterized as
\begin{equation} \tilde y_{\ell i}=S_{\ell}y_{\ell
i}\left({\lambda_{i}\over\lambda_{0}}\right)^{\alpha_{\ell}} \,\, .
\end{equation} The $y_{\ell i}$ values are the observed spectral
flux values of spectrum $\ell$.  The factor $S_{\ell}$ is a scale
factor and the factor
$\left({\lambda_{i}/\lambda_{0}}\right)^{\alpha_{\ell}}$ produces
tilt; $\lambda_{0}$ is just a reference wavelength that need not
actually be specified.  (This scaling and tilting of spectra is more
elaborate than the tilting we describe in the main text (see \S~1) for
spectra that we present and fit with \syn\ spectra.)

For two spectra labeled by indexes 1 and 2, we formulate the
$\chi^{2}$--like function:
\begin{eqnarray} \chi^{2}&=&\sum_{i}\left\{
\log\left[S_{2}y_{2i}\left({\lambda_{i}\over\lambda_{0}}\right)^{\alpha_{2}}\right]
-\log\left[S_{1}y_{1i}\left({\lambda_{i}\over\lambda_{0}}\right)^{\alpha_{1}}\right]
\right\}^{2} \cr &=&\sum_{i}\left\{\log\left[ \left({S_{2}\over
    S_{1}}\right) \left({y_{2i}\over y_{1i}}\right)
  \left({\lambda_{i}\over\lambda_{0}}\right)^{\alpha_{2}-\alpha_{1}}\right]\right\}^{2}
\cr &=&\sum_{i}\left(s+f_{i}+\alpha w_{i}\right)^{2} \cr
&=&\sum_{i}\left(s^{2}+f_{i}^{2}+\alpha^{2}w_{i}^{2} +2sf_{i}+2s\alpha
w_{i}+2\alpha f_{i}w_{i} \right) \cr
&=&s^{2}N+F_{f}+\alpha^{2}W_{w}+2sF+2s\alpha W+2\alpha F_{w} \,\, ,
\end{eqnarray}

\noindent where

\begin{equation}
s=\log\left({S_{2}\over S_{1}}\right) \,\, , , \qquad
f_{i}=\log\left({y_{2i}\over y_{1i}}\right)     \,\, ,  \qquad
w_{i}=\log\left({\lambda_{i}\over\lambda_{0}}\right)   \,\, , \qquad
\alpha=\alpha_{2}-\alpha_{1}                           \,\, ,
\end{equation}

\begin{equation}
N=\hbox{number of spectral points}   \,\, , \\
\end{equation}

\begin{equation}
F_{f}=\sum_{i}f_{i}^{2}         \,\, , \qquad
W_{w}=\sum_{i}w_{i}^{2}      \,\, ,   \qquad
F=\sum_{i}f_{i}              \,\, , \qquad
W=\sum_{i}w_{i}             \,\, ,
\end{equation}

\noindent and

\begin{equation}
F_{w}=\sum_{i}f_{i}w_{i}         \,\, .
\end{equation}

\noindent

     To minimize the $\chi^{2}$, we find the partial derivatives.

\begin{equation}
{\partial\chi^{2}\over\partial s}=2(sN+F+\alpha W)  \qquad\hbox{and}\qquad
{\partial\chi^{2}\over\partial \alpha}=2(\alpha W_{w}+sW+F_{w})  \,\, . 
\end{equation}

\noindent

Setting
these equations equal to zero and solving we find the $s$ and $\alpha$
values that give the stationary point of $\chi^{2}$: 

\begin{equation}
s={W_{w}F-WF_{w} \over W^{2}-W_{w}N}   \qquad\hbox{and}\qquad 
\alpha={WF-NF_{w} \over NW_{w}-W^{2}}   \,\,  .
\end{equation}

\noindent

\noindent Since the stationary point is unique and $\chi^{2}$ must
have an analytic global minimum with respect to $s$ and $\alpha$,
equation~(A8) gives the global minimizing values of $s$ and $\alpha$.

\clearpage     

\begin{figure}
\includegraphics[width=.8\textwidth,angle=270]{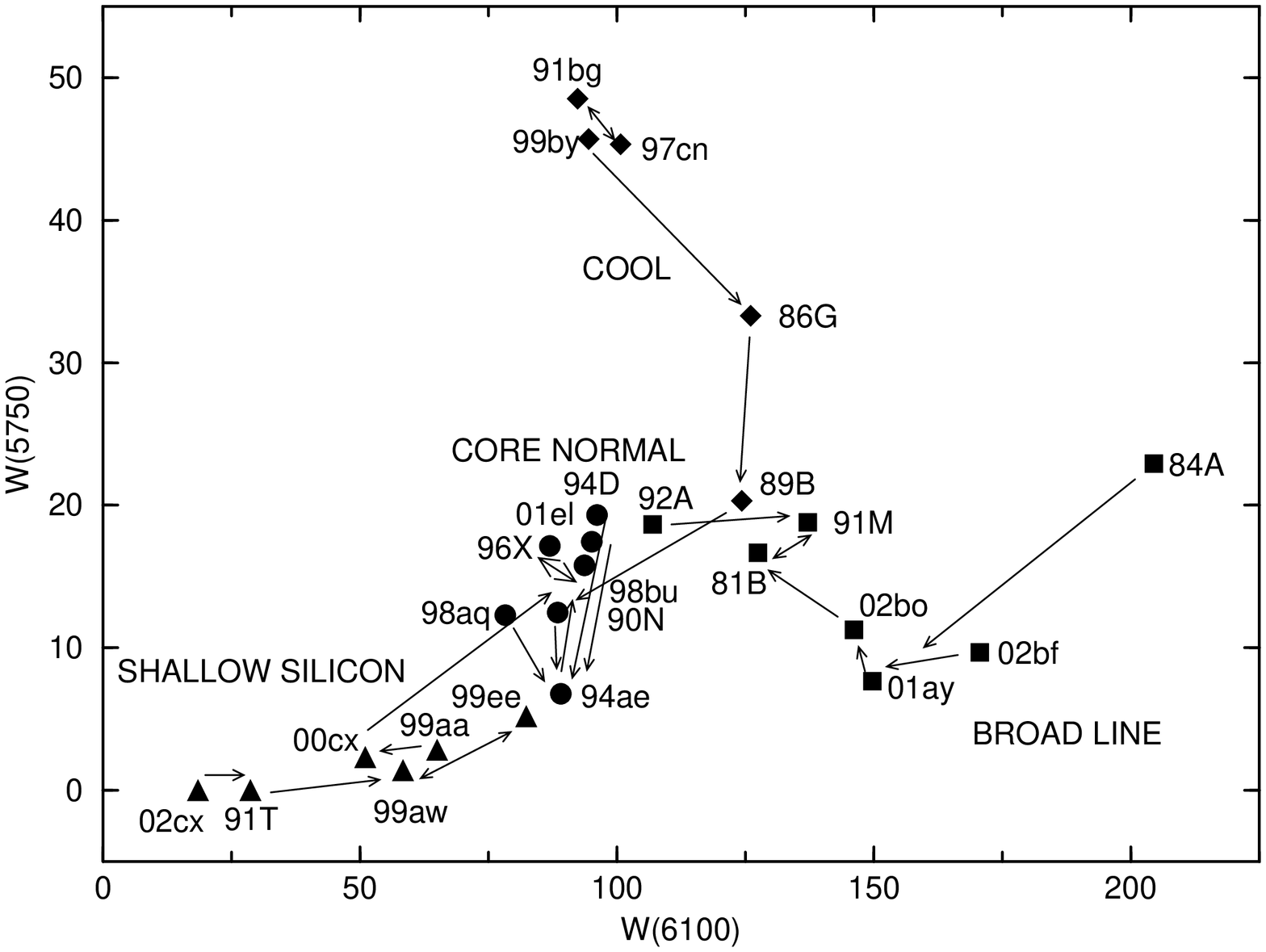}
\caption{\575\ is plotted against \610.  {\sl Single--headed arrows}
  point to nearest neighbors (see text). {\sl Double--headed arrows}
  connect mutual nearest neighbors.  Core--normal SNe~Ia are shown as
  {\sl circles}, broad--line SNe~Ia as {\sl squares}, cool SNe~Ia as
  {\sl diamonds}, and shallow--silicon SNe~Ia as {\sl triangles}.}
\end{figure}

\begin{figure}
\includegraphics[width=.8\textwidth,angle=0]{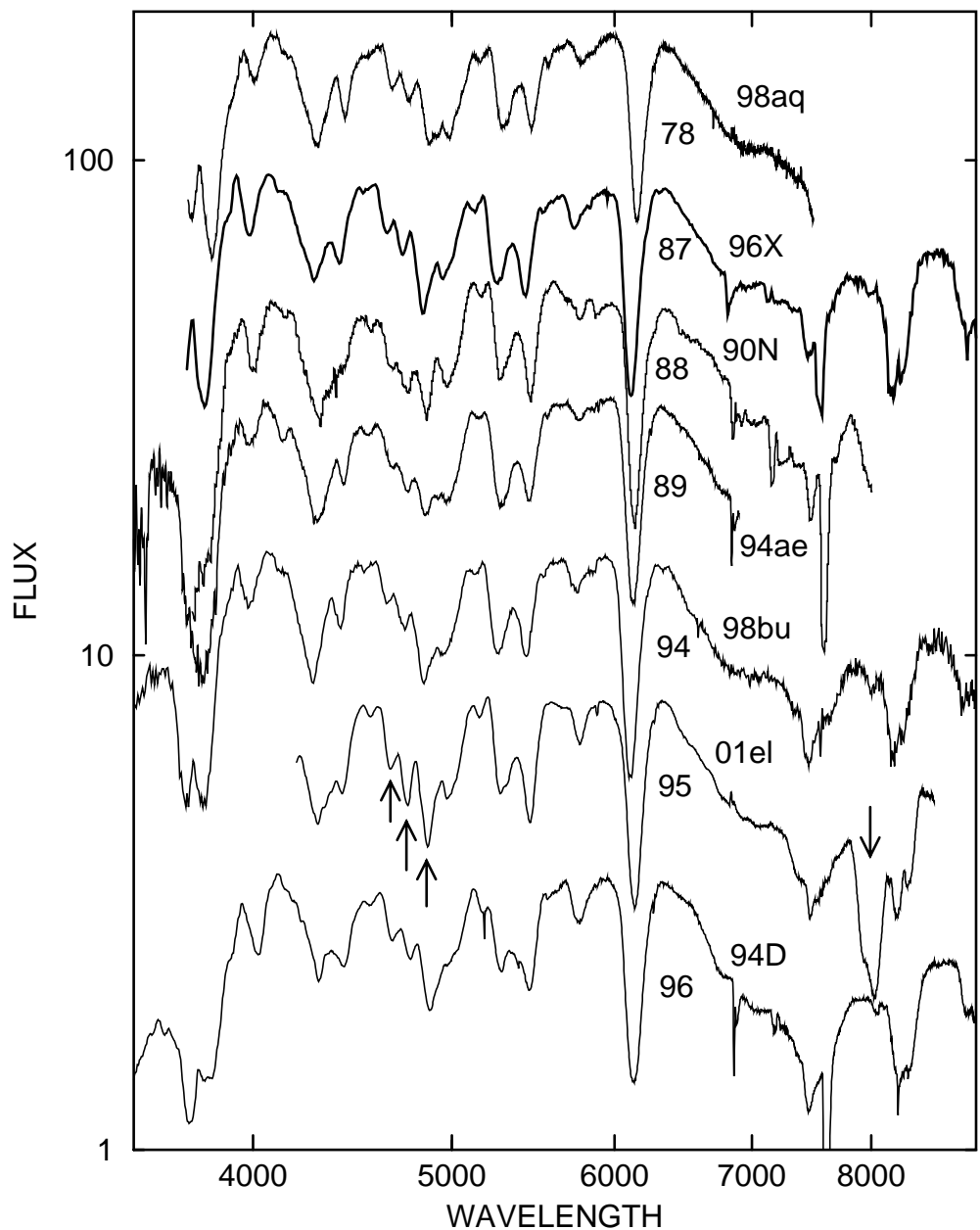}
\caption{Spectra of seven core--normals, ordered according to
increasing \610\ from top to bottom; the \610\ values are shown
adjacent to the 6100\ang\ absorptions.  The spectra have been tilted
to make the flux peaks near 4600\ang\ and 6300\ang\ equally high.
Vertical displacements are arbitrary.  Narrow absorptions near
7600\ang\ and 6900\ang\ are telluric.  The only conspicuous difference
among these spectra is the strong high--velocity Ca~II IR3 absorption
near 8000\ang\ in SN~2001el ({\sl downward arrow}).  Three less
conspicuous differences ({\sl upward arrows}), also in SN~2001el, are
attributed (see text) to high--velocity Fe~II.}  \end{figure}

\begin{figure}
\includegraphics[width=.8\textwidth,angle=270]{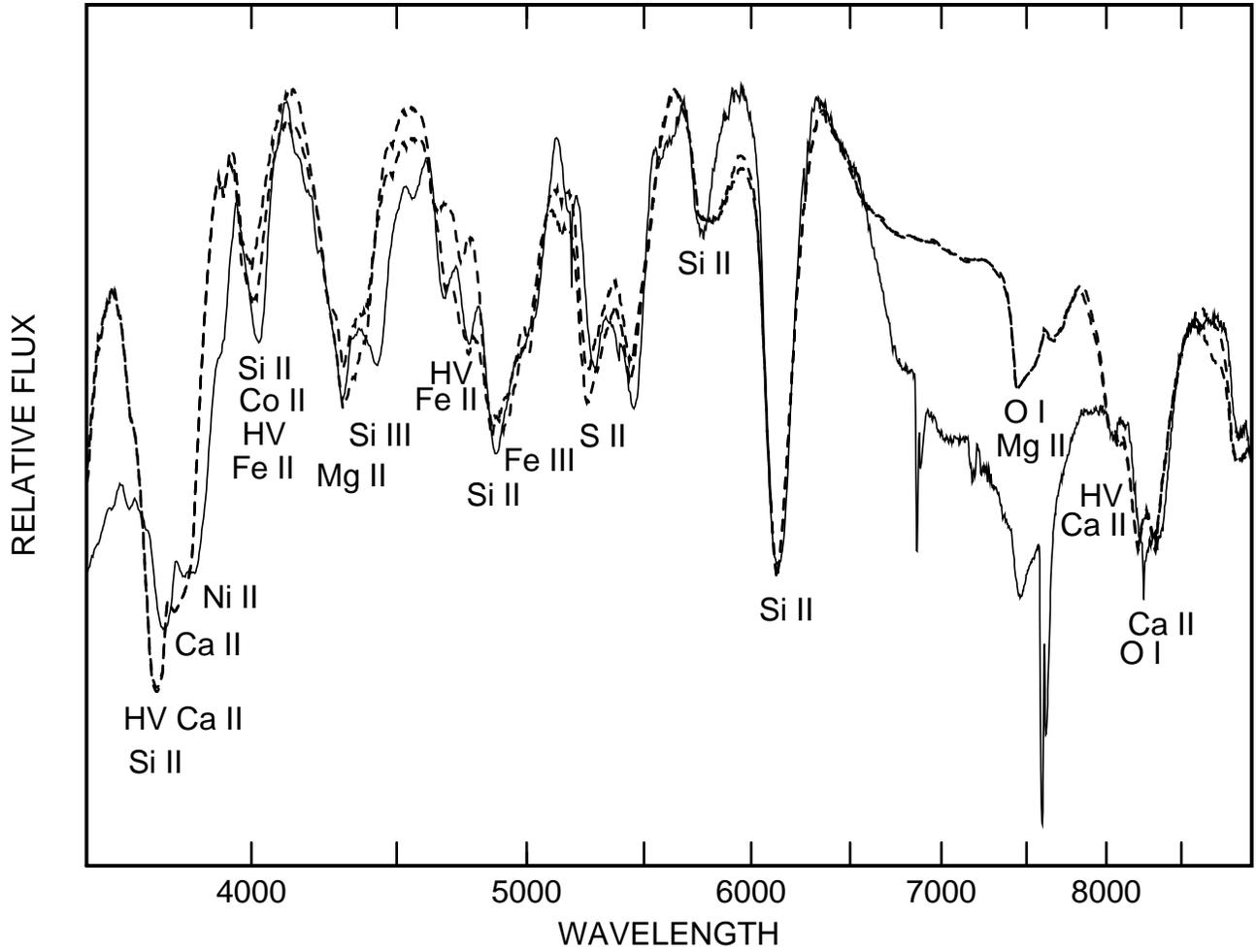}
\caption{The spectrum of SN~1994D ({\sl solid line}) is compared with
  two synthetic spectra ({\sl dashed lines}). One includes HV~Fe~II
  and the other does not.  The main difference in the synthetic
  spectra is due to HV~Fe~II absorption between 4600\ang\ and
  4800\ang.}
\end{figure}

\begin{figure}
\includegraphics[width=.8\textwidth,angle=270]{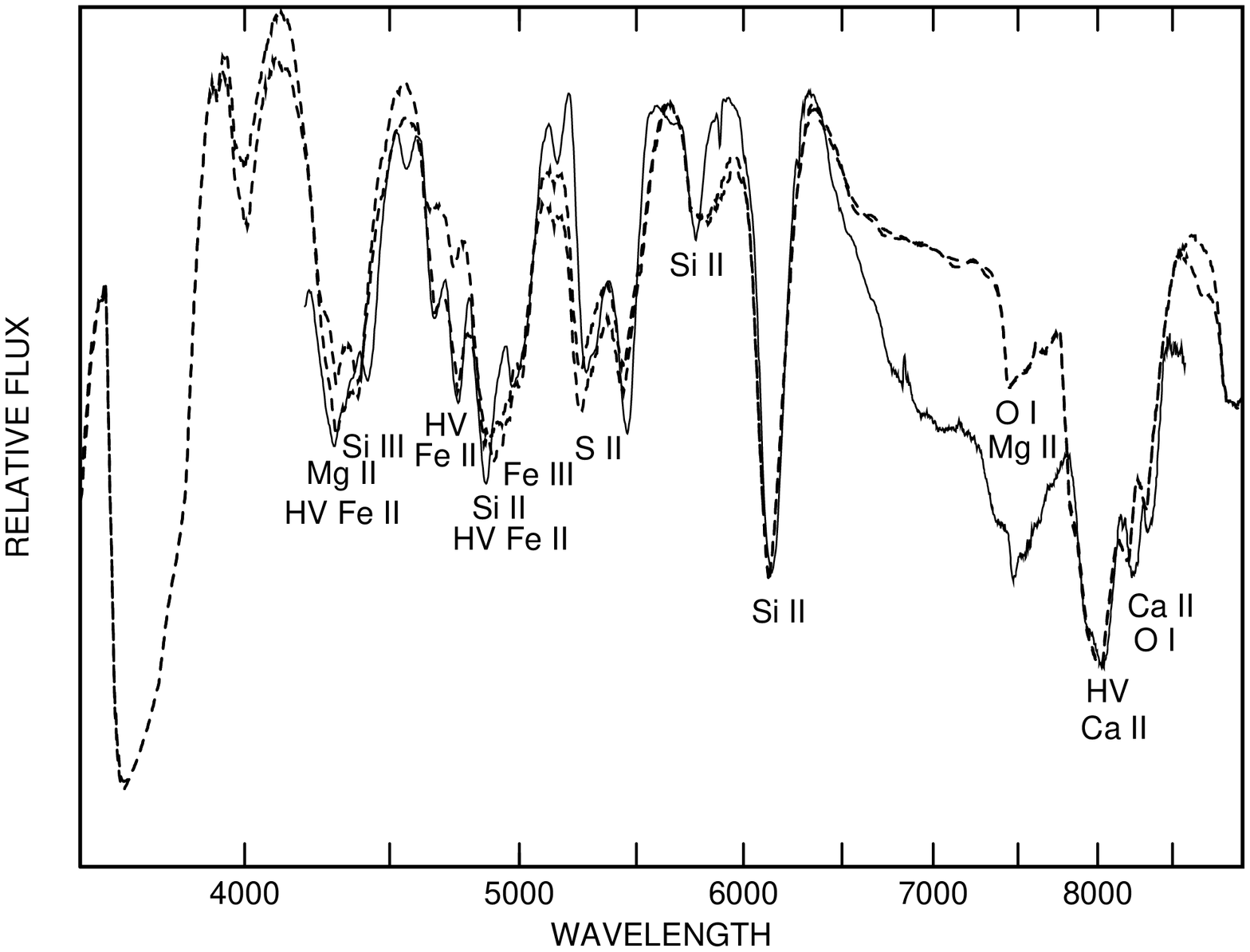}
\caption{The spectrum of SN~2001el ({\sl solid line}) is compared with
  synthetic spectra ({\sl dashed lines}) that do and do not include
  HV~Fe~II.}
\end{figure}

\begin{figure}
\includegraphics[width=.8\textwidth,angle=0]{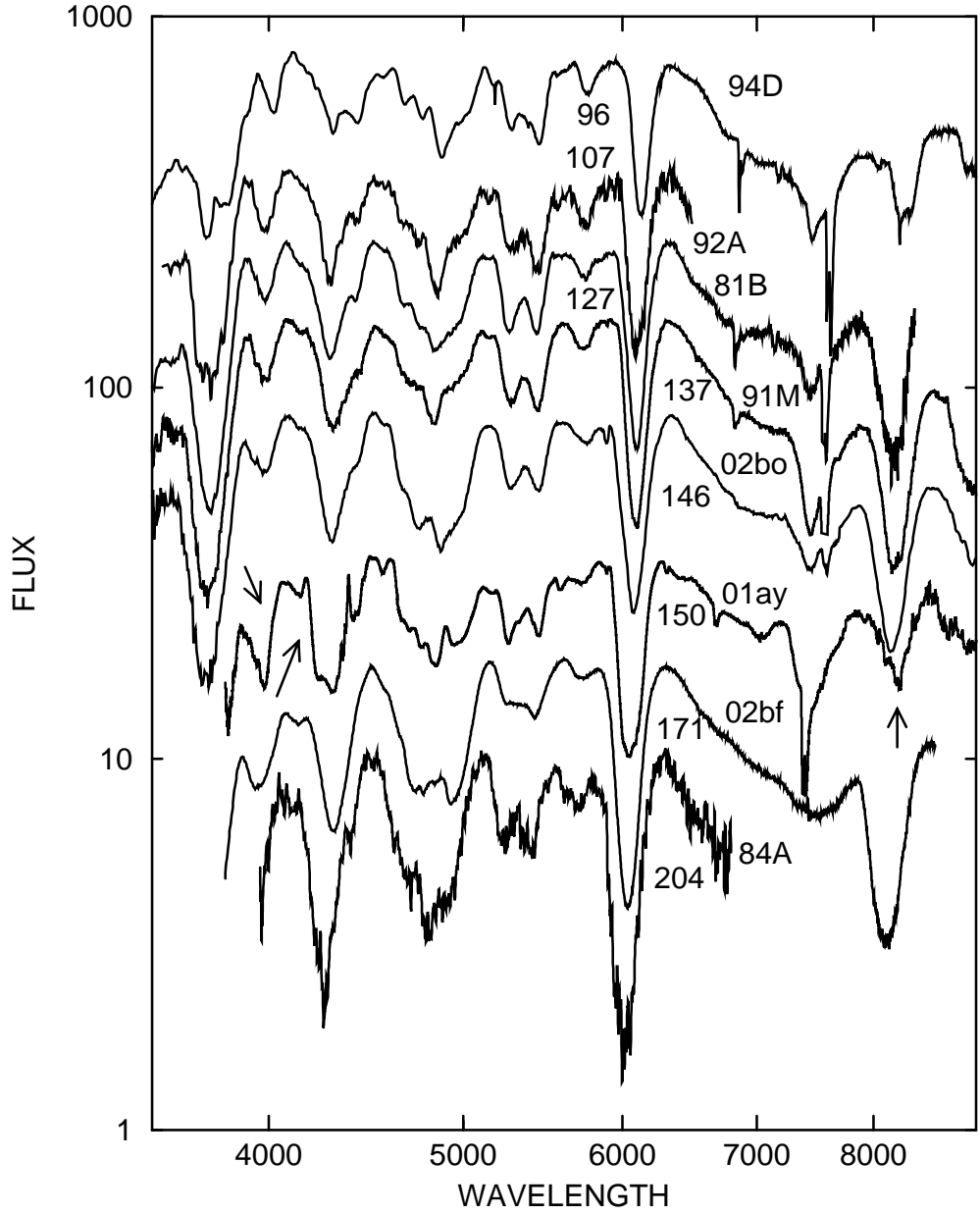}
\caption{Spectra of the core--normal SN~1994D and seven broad--line
  SNe~Ia. Three {\sl arrows} indicate some of the differences between
  SN~2001ay and the others.}
\end{figure}

\begin{figure}
\includegraphics[width=.8\textwidth,angle=270]{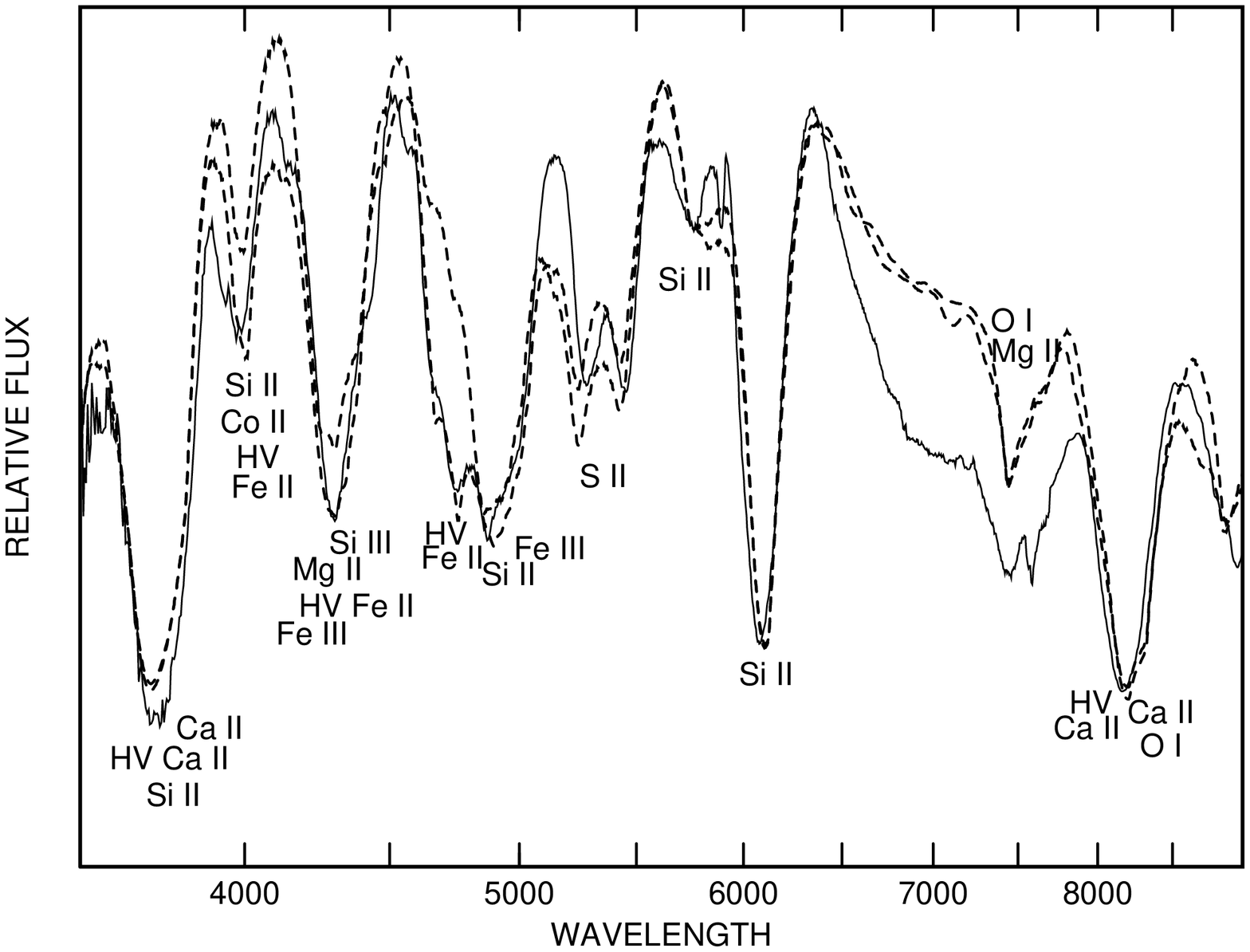}
\caption{The spectrum of SN~2002bo ({\sl solid line}) is compared with
synthetic spectra ({\sl dashed lines}) that do and do not include
HV~Fe~II. }
\end{figure}

\begin{figure}
\includegraphics[width=.8\textwidth,angle=270]{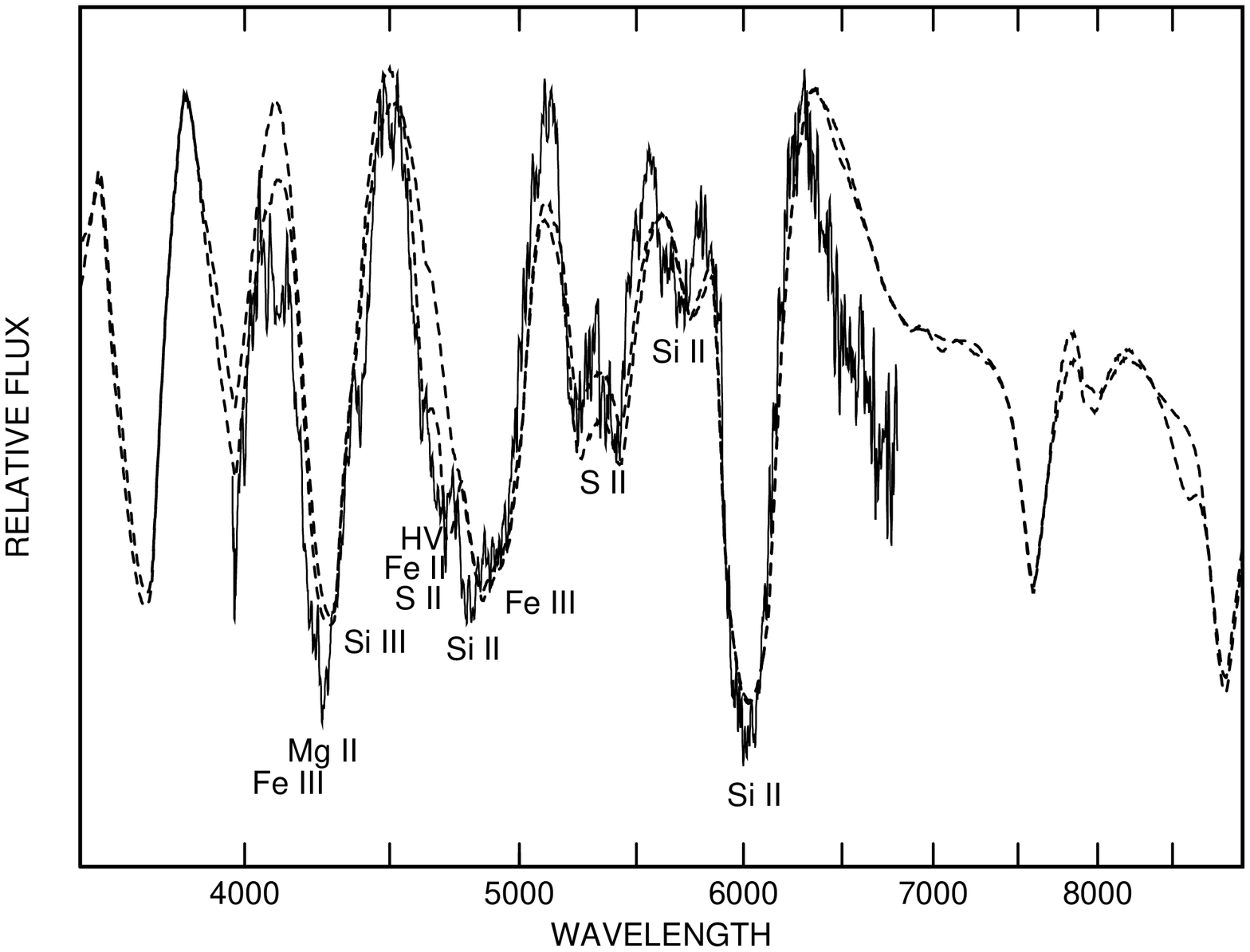}
\caption{The spectrum of SN~1984A ({\sl solid line}) is compared with
  synthetic spectra ({\sl dashed lines}) that do and do not include
  HV~Fe~II.}
\end{figure}

\begin{figure}
\includegraphics[width=.8\textwidth,angle=0]{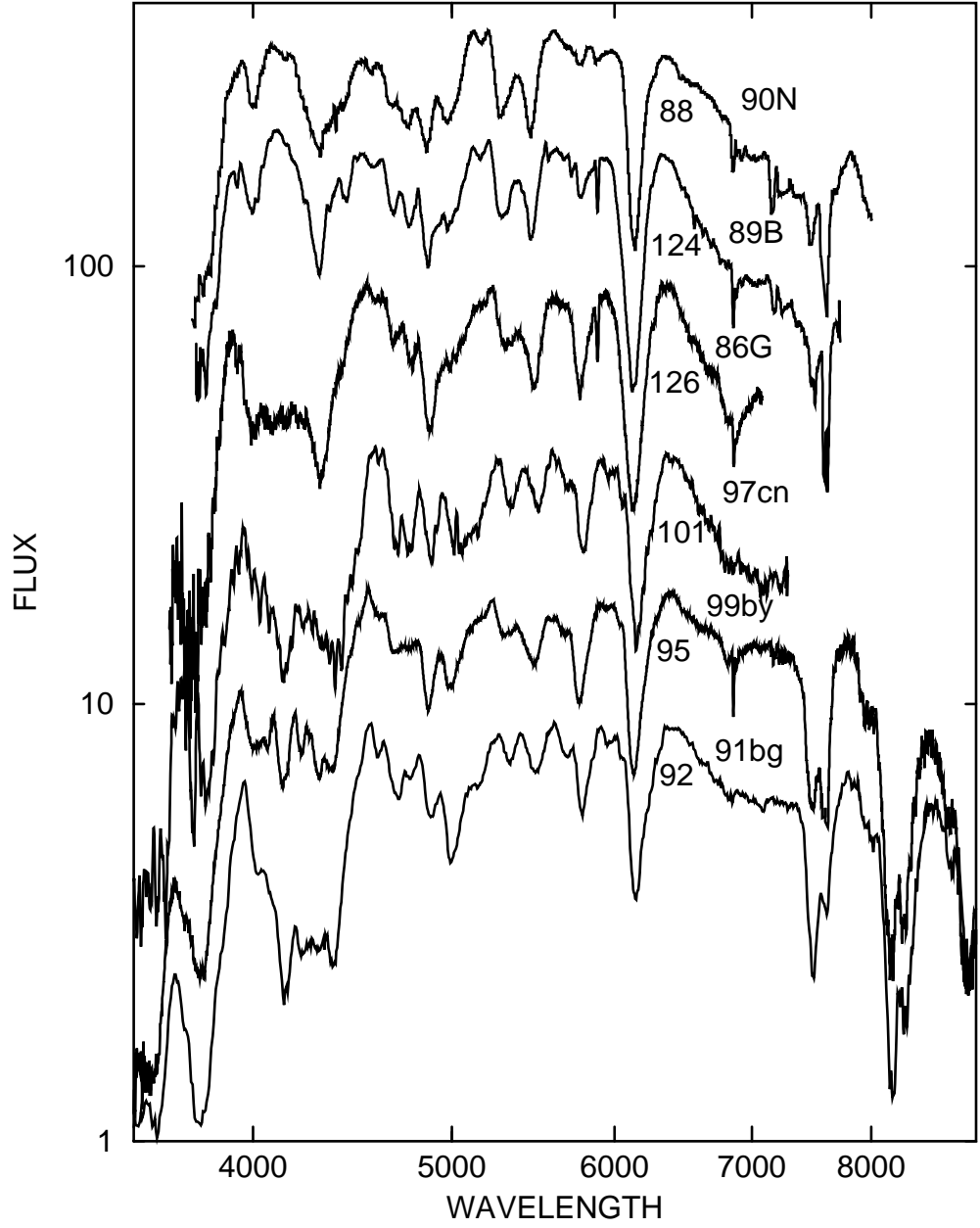}
\caption{Spectra of the core--normal SN~1990N and five cool SNe~Ia.}
\end{figure}

\begin{figure}
\includegraphics[width=.8\textwidth,angle=270]{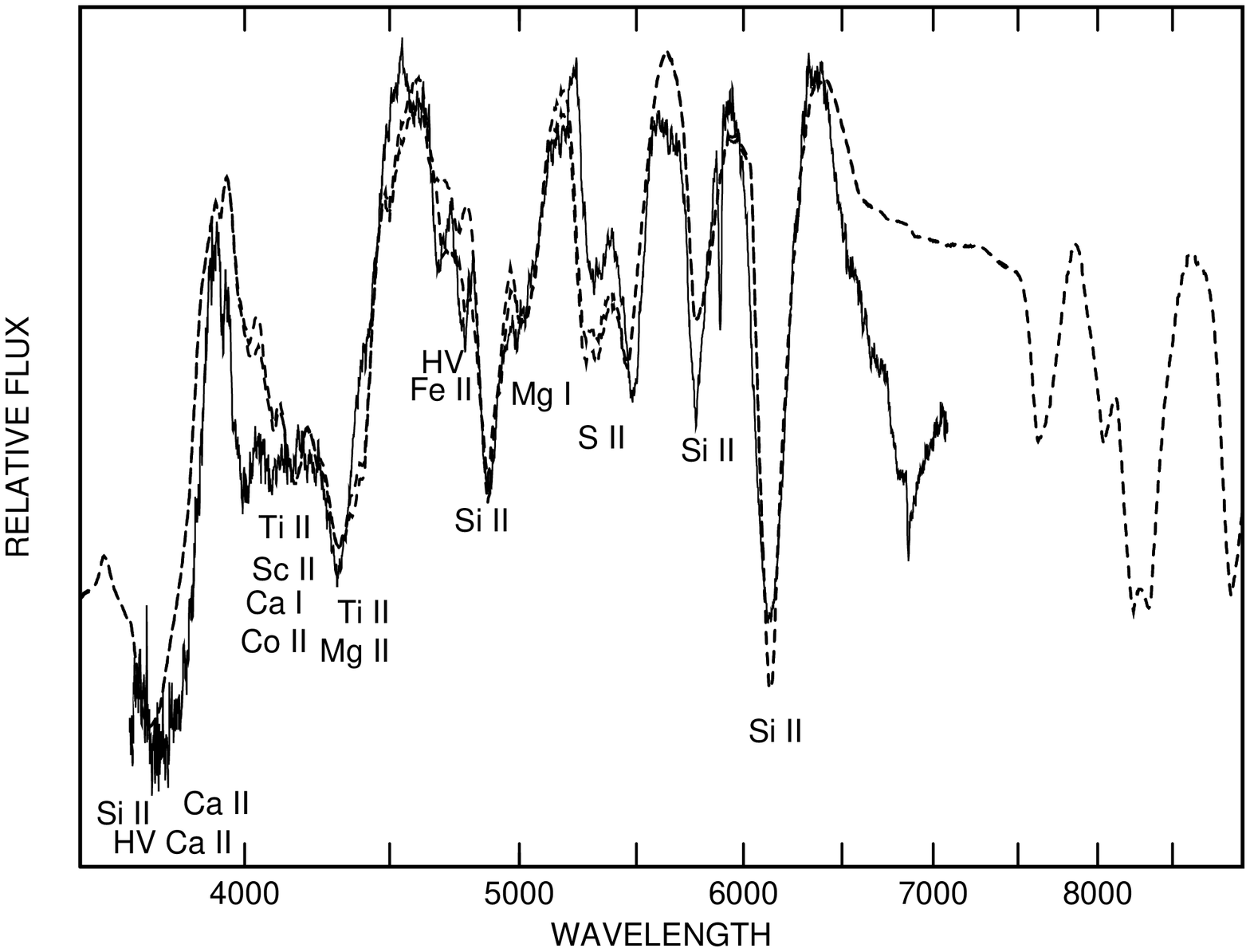}
\caption{The spectrum of SN~1986G ({\sl solid line}) is compared with
  synthetic spectra ({\sl dashed lines}) that do and do not include
  HV~Fe~II.}
\end{figure}

\begin{figure}
\includegraphics[width=.8\textwidth,angle=270]{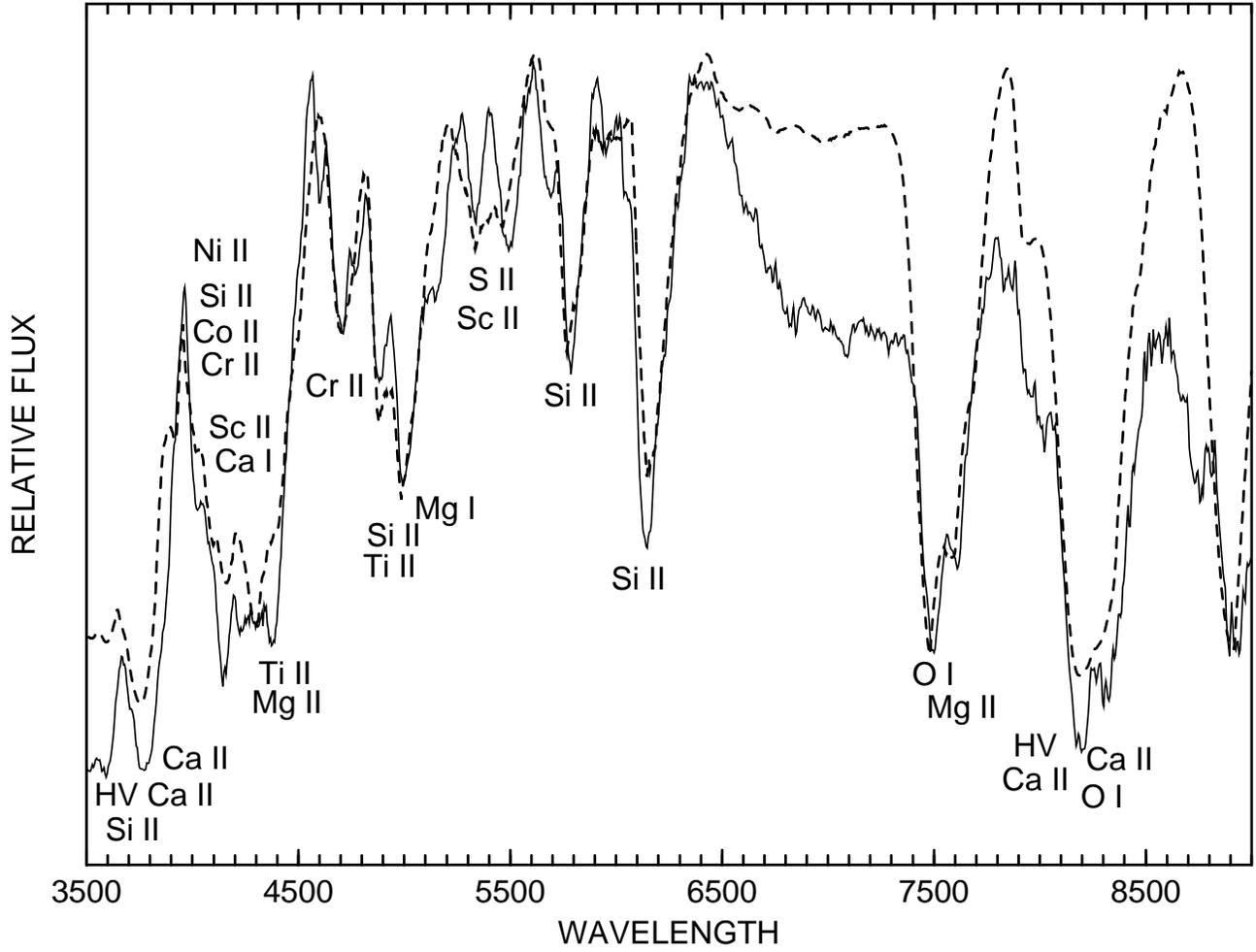}
\caption{The spectrum of SN~1991bg ({\sl solid line}) is compared with
  a synthetic spectrum ({\sl dashed line}).}
\end{figure}

\begin{figure}
\includegraphics[width=.8\textwidth,angle=0]{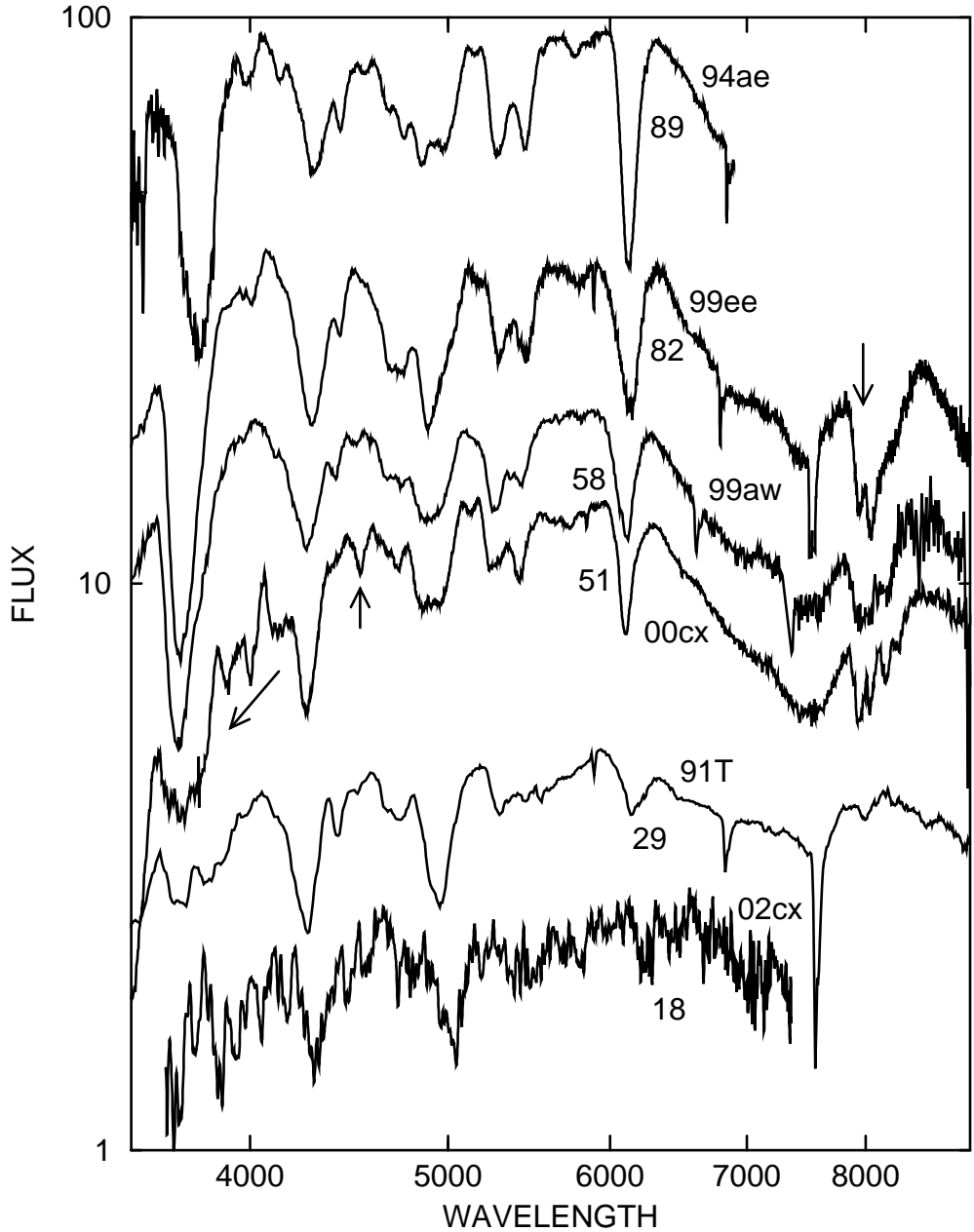}
\caption{Spectra of the core--normal SN~1994ae and five
shallow--silicon SNe~Ia. {\sl Arrows} indicate (from right to left)
the strong HV Ca~II IR3 absorption in SN~1999ee; the distinct
4530\ang\ absorption in SN~2000cx; and the depression of the spectrum
of SN~2000cx from about 3800\ang\ to 4200\ang.  } \end{figure}

\begin{figure}
\includegraphics[width=.8\textwidth,angle=270]{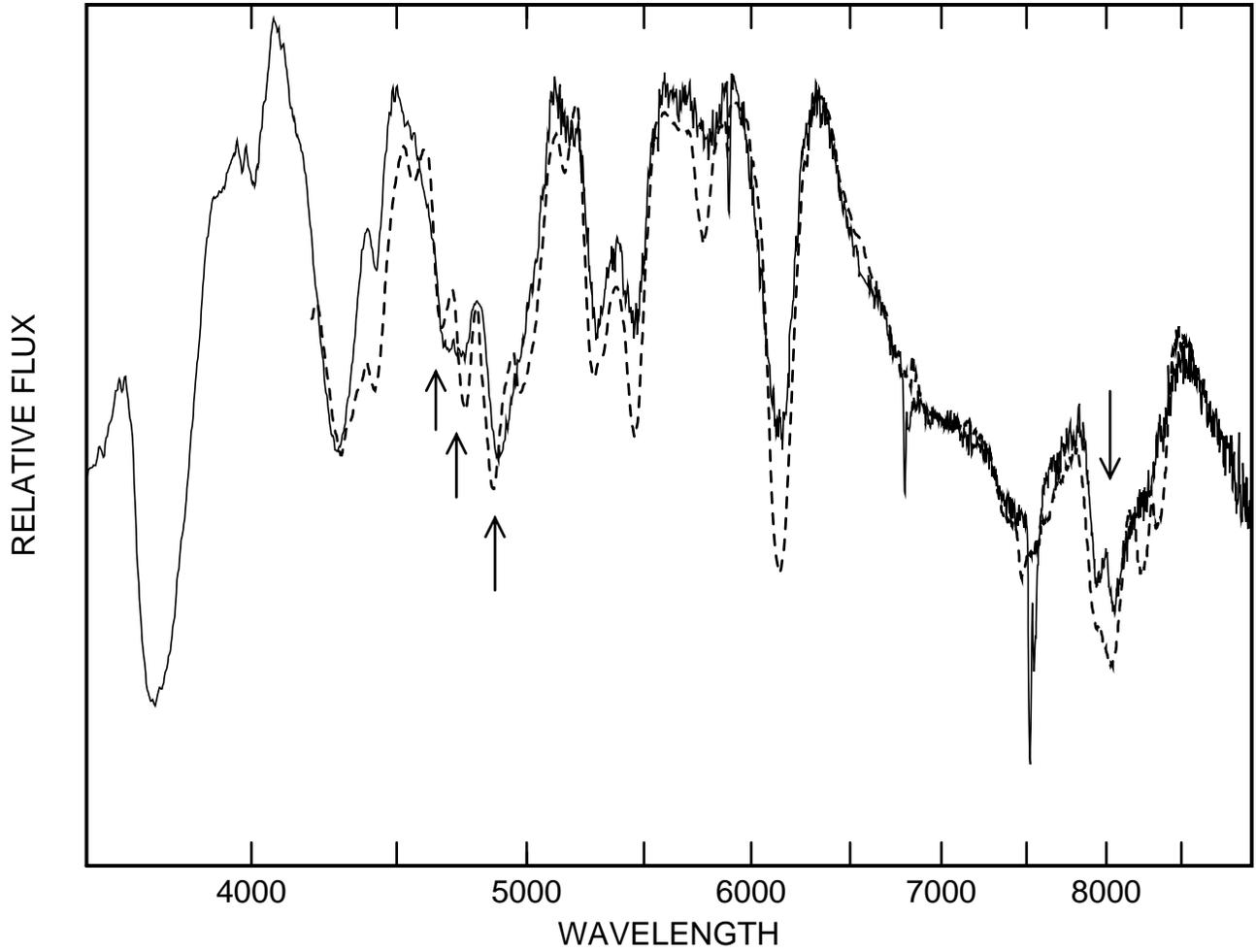}
\caption{The spectra of SN~1999ee ({\sl solid line}) and SN~2001el
({\sl dashed line}) are compared.  {\sl Arrows} indicate strong Ca~II
  IR3 absorption and enhanced HV~Fe~II absorption in both supernovae.}
\end{figure}

\begin{figure}
\includegraphics[width=.8\textwidth,angle=270]{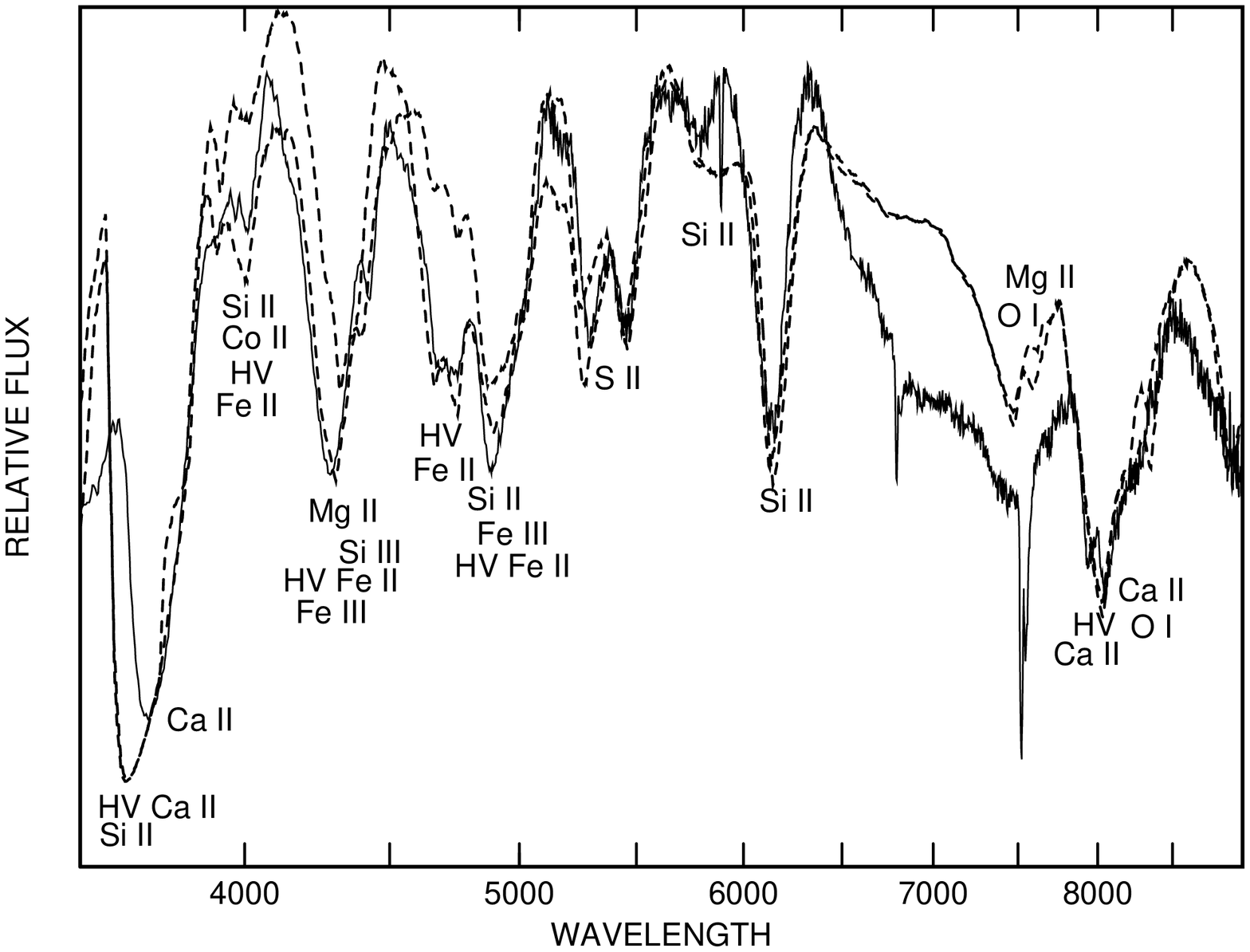}
\caption{The spectrum of SN~1999ee ({\sl solid line}) is compared with
  synthetic spectra ({\sl dashed lines}) that do and do not include
  HV~Fe~II.}
\end{figure}

\begin{figure}
\includegraphics[width=.8\textwidth,angle=270]{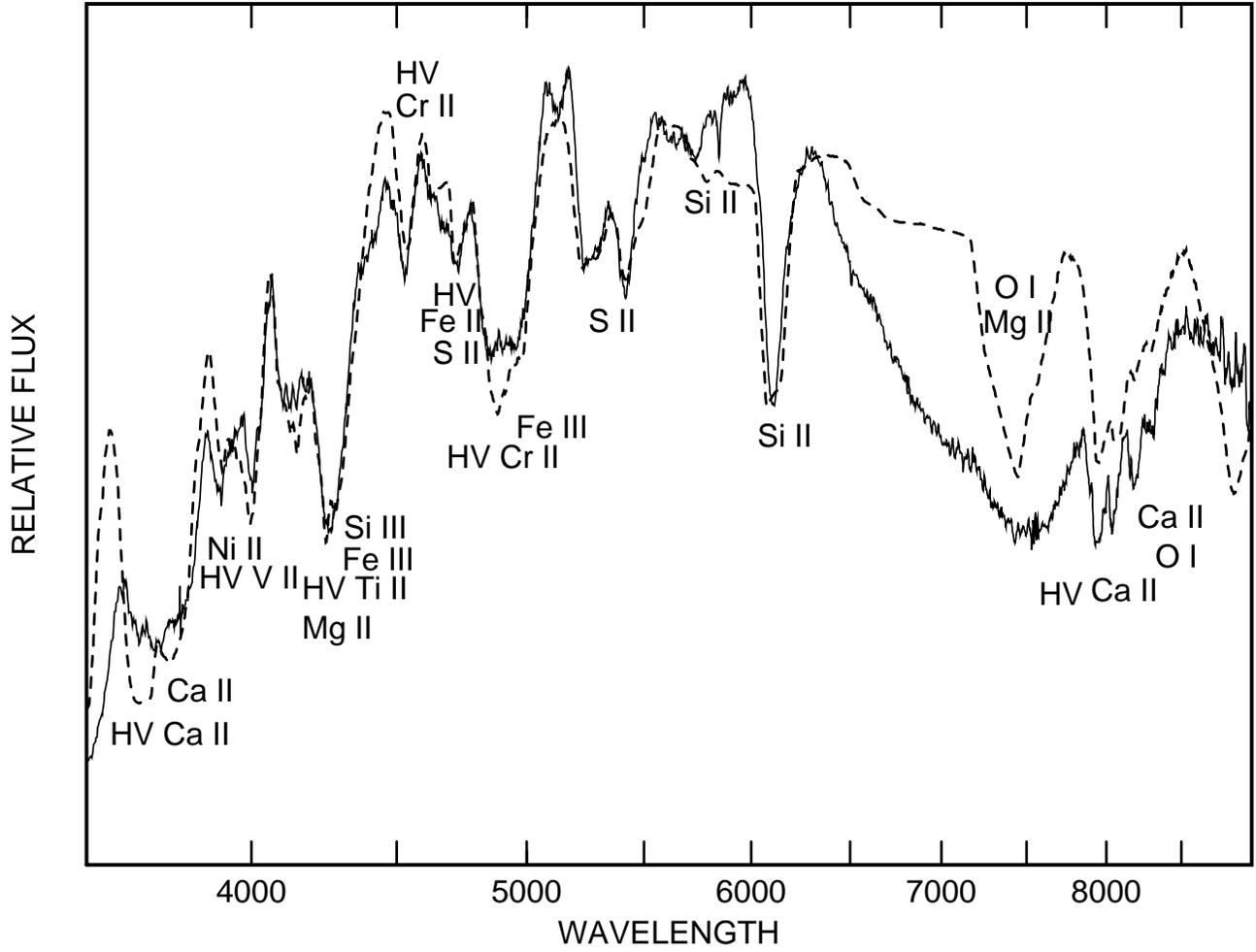}
\caption{The spectrum of SN~2000cx ({\sl solid line}) is compared with
  a synthetic spectrum ({\sl dashed line}).}
\end{figure}


\clearpage

\begin{deluxetable}{lccccl}
\tablenum{1}
\setlength{\tabcolsep}{4pt}

\tablecaption{The SN Ia Sample}

\tablehead{ \colhead{SN} & \colhead{epoch} & \colhead{galaxy} &
\colhead{\610} & \colhead{\575} & \colhead{reference} \\

\colhead{} & \colhead{(days)} & \colhead{} &
\colhead{(\ang)} & \colhead{(\ang)} & \colhead{} }

\startdata

1981B & $-$2 & NGC 4536 & 127 & 17 & Branch et al. (1983)\\

1984A & $-$3 & NGC 4419 & 204 & 23 & Barbon et al. (1989)\\

1986G & $-$1 & NGC 5128 & 126 & 33 & Cristiani et
al. (1992)\\

1989B & 0 & NGC 3627 & 124 & 20 & Wells et al. (1994)\\

1990N & $-$2 & NGC 4639 & 88 & 12 & Asiago SN Group\\

1991M & 3 & IC 1151 & 137 & 19 & Gomez et~al. (1996)\\

1991T & $-$3 & NGC 4527 & 29 & 0 & Phillips et~al. (1992)\\

1991bg & 0 & NGC 4374 & 92 & 49 & Filippenko et~al. (1992a)\\

1992A & $-$1 & NGC 1380 & 107 & 19 & P.~Challis, unpublished\\

1994D & $-$1 & NGC 4526 & 96 & 19 & Meikle et~al. (1996)\\

1994ae & 0 & NGC 3370 & 89 & 7 & Howell \& Nugent (2004)\\

1996X & $-$2 & NGC 5061 & 87 & 17 & Salvo et al. (2001)\\

1997cn & 3 & NGC 5490 & 101 & 45 & Turatto et~al. (1998)\\

1998aq & 0 & NGC 3982 & 78 & 12 & Branch et al. (2003)\\

1998bu & $-$1 & NGC 3368 & 94 & 16 & Hernandez et~al. (2000)\\

1999aw & 3 & $---$ & 58 & 1 & Howell \& Nugent (2004)\\

1999by & $-$3 & NGC 2841 & 95 & 46 & Garnavich et al. (2004)\\

1999ee & $-$2 & IC 5179 & 82 & 5 & Hamuy et al. (2002)\\

2000cx & 2 & NGC 524 & 51 & 2 & Li et al. (2002)\\

2001ay & 0 & IC 4423 & 150 & 8 & Howell \& Nugent (2004)\\

2001el& 1 & NGC 1448 & 95 & 16 & Wang et al. (2003)\\

2002bf & 3 & $---$ & 171 & 10 & Leonard et al. (2005)\\

2002bo & $-$1 & NGC 3190 & 146 & 11 & Benetti et al. (2004)\\

2002cx & $-$1 & $-$$-$$-$ & 18 & 0 & Li et al. (2003)\\
\enddata

\end{deluxetable}

\begin{deluxetable}{lccccccc}
\rotate
\tablenum{2}
\setlength{\tabcolsep}{4pt}

\tablecaption{Fitting Parameters for Core--Normals}

\tablehead{\colhead{ } & \colhead{SN~1998aq} & \colhead{SN~1996X} &
\colhead{SN~1990N} & \colhead{SN~1994ae} & \colhead{SN~1998bu} &
\colhead{SN~2001el} & \colhead{SN~1994D}}


\startdata

\vphot\ (\kms) & 12,000 & 13,000 & 12,000 & 12,000 & 13,000 & 12,000 &
12,000\\

\ve\ (\kms)& 1000 & 1000 & 1000 & 1000 & 1000 & 1000 & 1000\\

\tex\ (K) & 10,000 & 10,000 & 10,000 & 10,000 & 10,000 & 10,000 &
10,000\\

$\tau$(O~I) & & 0.7/14 & 0.5/14 &  & 0.7/14 & 0.5/14 & 0.5/14\\

$\tau$(Mg~II) & 1.5 & 1.2 & 1.7 & 1.3 & 2 & 1.5 & 1.2\\

$\tau$(Si~II) & 6 & 13 & 11 & 8 & 13 & 12 & 12\\

$\tau$(Si~III) & 1.7 & 0.8 & 1 & 1.2 & 0.6 & 1.7 & 0.6\\

$\tau$(S~II) & 2 & 2.5 & 2.3 & 2.4 & 2.3 & 2.5 & 2.3\\

$\tau$(Ca~II) & 80 & 80 & 80 & 80 & 80 & 120 & 80\\

$\tau$(HV Ca~II) &  & 2(2)/21 &  &  & 5(2)/21 & 12(12)/19[30] & 3(2)/20\\

$\tau$(HV Fe~II) & 0.5(2)/16 & 0.3(2)/18 & 0.6(2)/17 & 0.5(2)/17 &
0.3(2)/18 & 0.8(2)/17 & 0.7(2)/16\\

$\tau$(Fe~III) & 1.5 & 1.5 & 0.8 & 1.2 & 1.5 & 1.2 & 0.8\\

$\tau$(Co~II) & 0.6 & 0.6 & 0.4 & 0 & 0.6 &  & 0.8\\

$\tau$(Ni~II) & 0.7 & 0.2 & 0.8 & 0.5 & 0.5 & & 0.6\\
\enddata

\tablenotetext{a}{For each ion the optical depth, $\tau$, is the
optical depth at the photosphere or the detachment velocity of the
ion's reference line (ordinarily the ion's strongest line in the
optical spectrum).  When a value of \ve\ other than the default value
in row~2 is used, the value (in units of 1000~\kms) is given in
parentheses.  Minimum and maximum velocities (in units of 1000~\kms)
are preceded by a forward slash.  Maximum velocities are in square
brackets.  The absence of an optical--depth entry means that the
observed spectrum does not include the relevant wavelength range.}

\end{deluxetable}

\begin{deluxetable}{lccccccc}
\rotate
\tablenum{3}
\setlength{\tabcolsep}{4pt}

\tablecaption{Fitting Parameters for Broad--Line SNe~Ia}

\tablehead{\colhead{ } & \colhead{SN~1992A} & \colhead{SN~1981B} &
\colhead{SN~1991M} & \colhead{SN~2002bo} & \colhead{SN~2001ay} &
\colhead{SN~2002bf} & \colhead{SN~1984A}}

\startdata

\vphot\ (\kms)& 12,000 & 12,000 & 12,000 & 12,000 & 12,000 & 12,000 &
12,000\\

\ve\ (\kms)& 2000 & 2000 & 2000 & 2000 & 2000 & 2000 & 2000\\

\tex\ (K)& 10,000 & 10,000 & 10,000 & 10,000 & 10,000 & 10,000 &
10,000\\

$\tau$(O~I) & 0.5/14 & 0.5/14 & 0.5/14 & 0.5/14 & 0.5(4) & 0.3(10) & 0\\

$\tau$(Mg~II) & 1.2 & 1 & 1 & 1 & 1(4) & 2(4) & 5\\

$\tau$(Si~II) & 8 & 8 & 8 & 8 & 3(4) & 4(5)/[22] & 6(5)/[25]\\

$\tau$(Si~III) & 0.4 & 0.6 & 0.6 & 0.0 & 0.4 & 0.6 & 0.6\\

$\tau$(S~II) & 1.5 & 1.5 & 1.5 & 1.5 & 1 & 0.7 & 1.5\\

$\tau$(Ca~II) & 100 & 60 & 100 & 120 & 12 & 150 & \\

$\tau$(HV Ca~II) & 2/21 & 5/21 & 2/21 & 3(3)/21 & 0 &
40(3)/18[30] & \\

$\tau$(HV Fe~II) & 0.6/17 & 0.6/17 & 0.6/17 & 1.5(3)/17 &
0.3/17 & 2(3)/17 & 1(3)/20\\

$\tau$(Fe~III) & 0.5 & 1 & 0.2 & 1.2 & 0.3 & 1.2 & 2\\

$\tau$(Co~II) &1 & 1 & 1 & 1 & 1 &1  & \\

$\tau$(Ni~II) & 0.2 & 0.2 & 0.2 & 0.2 &  & 0.2&\\
\enddata
\end{deluxetable}

\begin{deluxetable}{lccccc}
\rotate
\tablenum{4}
\setlength{\tabcolsep}{4pt}

\tablecaption{Fitting Parameters for Cool SNe~Ia}

\tablehead{\colhead{ } & \colhead{SN~1989B} & \colhead{SN~1986G} &
\colhead{SN~1997cn} & \colhead{SN~1999by} & \colhead{SN~1991bg} }

\startdata

\vphot\ (\kms) & 11,000 & 11,000 & 11,000 & 11,000 & 11,000 \\

\ve\ (\kms)& 1000 & 1000 & 1000 & 1000 & 1000 \\

\tex\ (K) & 10,000 & 7000 & 7000 & 7000 & 7000 \\

$\tau$(O~I) &0.5/12 & &  & 12/12 & 18/12 \\

$\tau$(Na~I) & 0 & 0 & 0.3/12 & 0.5/12 & 0.3/12 \\

$\tau$(Mg~I) & 0 & 3 & 2/12 & 3.5/12 & 7/12\\

$\tau$(Mg~II) & 4 & 8 & 20/12 & 20/12 & 30/12\\

$\tau$(Si~II) & 15 & 100/[15] & 300/[13] & 150/[14] & 200/[13]\\

$\tau$(Si~III) & 0.6 & 0 & 0 & 0 & 0\\

$\tau$(S~II) & 2.3 & 2 & 0.8 & 1 & 0.5 \\

$\tau$(Ca~I) & 0 & 8 & 10 & 10 & 10 \\

$\tau$(Ca~II) & 120 & 500 & 5000 & 5000 & 5000 \\

$\tau$(HV Ca~II) & & 10(2)/19 & 50/16& 50/16 & 50/16 \\

$\tau$(Sc~II) & 0 & 8 & 10 & 10 & 10 \\

$\tau$(Ti~II) & 0 & 2.5 & 5 & 4 & 4 \\

$\tau$(Cr~II) & 0 & 0 & 2 & 2 & 2 \\

$\tau$(HV Fe~II) & 0.6(2)/17 & 1/15 & 1/17 & 0 & 0\\

$\tau$(Fe~III) & 1 & 0 & 0 & 0 & 0 \\

$\tau$(Co~II) & 0.4 & 3 & 03 & 4 &3 \\

$\tau$(Ni~II) & 0.8 & 0.8 & 3 & 4 & 3\\
\enddata
\end{deluxetable}

\begin{deluxetable}{lcccccc}
\rotate
\tablenum{5}
\setlength{\tabcolsep}{4pt}

\tablecaption{Fitting Parameters for Shallow--Silicon SNe~Ia}

\tablehead{\colhead{ } & \colhead{SN~1999ee} & \colhead{SN~1999aw} &
\colhead{SN~2000cx} & \colhead{SN~1991T} & \colhead{SN~2002cx}}

\startdata

\vphot\ (\kms)& 11,000 &  13,000 & 13,000 & 11,000 & 6000 \\

\ve\ (\kms)& 1000 &  1000 & 1000 & 1000 & 1000 \\

\tex\ (K)& 10,000 &  10,000 & 10,000 & 10,000 & 10,000 \\

$\tau$(O~I) &0.3(6)/12 & 0.3(6) & 0.5(6)/[25] & 0.4 & \\

$\tau$(Mg~II) & 2 &  0.5(4) & 1.5 & 0 & 0.4\\

$\tau$(Si~II) & 4 &  1.5/15 & 3/15[16] & 0.4 & 0.6\\

$\tau$(Si~III) & 1 & 0.5 & 0.5/14 & 1.5 & 0.8\\

$\tau$(S~II) & 2 &  1.6 & 0.8/14 & 0.4 & 0.6\\

$\tau$(Ca~II) & 70 & 30 & 100 & 1.8 & 3 \\

$\tau$(HV Ca~II) & 10(6)/19 &  8(6)/19& 80/23 & 0.4(2)/20 &
0.5/14\\

$\tau$(HV Sc~II) & 0 &0  & 0 & 0 &0  \\

$\tau$(HV Ti~II) & 0  &0  & 1/23 & 0 &0  \\

$\tau$(HV V~II) & 0  &0  & 4/23 & 0 &0  \\

$\tau$(HV Cr~II) & 0 &0  & 1/22 & 0 &0  \\

$\tau$(HV Fe~II) & 0.6(4)/18 & 0.2(4)/18 & 0.8(5)/20 &
0.15(2)/19 & 0.4/13 \\

$\tau$(Fe~III) & 0.8 & 0.8 & 0.8 & 0.9 & 0.8\\

$\tau$(Co~II) & 0.8 & 0.2 & 0 & 0 & 0 \\

$\tau$(Ni~II) & 1.8 & 1.2 & 1.5 & 0 &0.4\\
\enddata
\end{deluxetable}
\end{document}